\documentclass[useAMS,usenatbib]{mn2e}
\usepackage{graphicx}

\newcommand{\swift}{{\it Swift~\/}}

\newcommand{\suzaku}{{\it Suzaku~\/}}
\newcommand{\fermi}{{\it Fermi~\/}}

\def\H0{{\rm ~km~s^{-1}~Mpc^{-1}}}

\def\la{\mathrel{\hbox{\rlap{\hbox{\lower4pt\hbox{$\sim$}}}{\raise2pt\hbox{$<$}}}}}
\def\ga{\mathrel{\hbox{\rlap{\hbox{\lower4pt\hbox{$\sim$}}}{\raise2pt\hbox{$>$}}}}}
\def\d25{D$_{25}$}

\def\1007{GRB~061007\/}

\title[Spectral components in GRB~061007] {Spectral components in the
  bright, long GRB~061007: properties of the photosphere and the nature of the outflow}

\author[J.~Larsson, et al.]  {\parbox[]{6.0in}
  {J.~Larsson$^{1,2}$\thanks{E-mail:josefin.larsson@astro.su.se },
    F.~Ryde$^{3,2}$, C.~Lundman$^{3,2}$, S.~McGlynn$^{3,2}$, S.~Larsson$^{1,3,2}$, M.~Ohno$^4$,
    K.~Yamaoka$^5$\\ 
\footnotesize {\it $^1$Department of Astronomy, Stockholm University, SE-106 91 Stockholm, 
Sweden 
\\ $^2$The Oskar Klein Centre for Cosmoparticle Physics, AlbaNova, SE-106 
91 Stockholm, Sweden
\\$^3$Department of Physics, Royal Institute of Technology(KTH), AlbaNova, 
SE-106 91 Stockholm, Sweden
\\$^4$Institute of Space and Astronautical Science, Japan Aerospace Exploration Agency,
3-1-1 Yoshinodai, Chuo-ku Sagamihara, Kanagawa 252-5120, Japan
\\$^5$Department of Physics and Mathematics, Aoyama Gakuin University, 
5-10-1 Fuchinobe, Chuo-ku, Sagamihara, Kanagawa 252-5258, Japan  
}}}

\date{Accepted 2011 February 21. Received 2011 February 15; in original form 2011 January 18}

\pagerange{\pageref{firstpage}--\pageref{lastpage}}
\pubyear{2010}

\begin{document}

\maketitle

\label{firstpage}

\begin{abstract}
We present a time-resolved spectral analysis of the bright, long
GRB~061007 ($z=1.261$) using \swift BAT and \suzaku WAM data. We find
that the prompt emission of GRB~061007 can be equally well explained
by a photospheric component together with a power law as by a Band
function, and we explore the implications of the former model. The
photospheric component, which we model with a multicolour blackbody,
dominates the emission and has a very stable shape throughout the
burst. This component provides a natural explanation for the
hardness-intensity correlation seen within the burst and also allows
us to estimate the bulk Lorentz factor and the radius of the
photosphere. The power-law component dominates the fit at high
energies and has a nearly constant slope of $-1.5$. We discuss the
possibility that this component is of the same origin as the
high-energy power laws recently observed in some \fermi LAT bursts.

 \end{abstract}
            

\section{Introduction}
\label{introduction}

The prompt emission of gamma-ray bursts (GRBs) is usually well
modelled by a smoothly broken power law (\citealt{Band93}), peaking in
the 100--1000~keV energy range. The physical origin of this
emission is still unclear, and the situation has recently become even
more puzzling, with the \fermi satellite revealing that at least some
bright bursts have additional power-law components extending well into
the GeV energy range (\citealt{Abdo09,Ackermann10,Granot10}).  For a
long time synchrotron emission has been considered the most promising
candidate for explaining the Band component of the spectra. Although
this model can explain many aspects of the emission, it also suffers
from some significant problems. In particular, many bursts are
observed to have harder low-energy spectra than predicted by standard
synchrotron models (\citealt{Crider97,Preece98,Ghirlanda03}).

An alternative model that is able to account for the hard low-energy
spectra is photospheric emission
(\citealt{Meszaros00,Meszaros02,Rees05,Peer06}). Such a model also has
the advantage of providing a natural explanation for the observed
correlations between the peak energy and luminosity within/between
bursts (\citealt{Golenetskii83,Borgonovo01,Yonetoku04}). Fits to
observational data also support this picture, with blackbody models
(often in combination with an additional non-thermal component)
providing excellent descriptions of many bursts
(\citealt{Ghirlanda03,Ryde04,Ryde09}).  However, in spite of its
successes, it seems clear that single, narrow blackbody components do
not significantly contribute to the prompt emission of most bursts
(\citealt{Ghirlanda07,Bellm10}).

This does not necessarily rule out a photospheric origin for the
emission.  In fact, there are many reasons to expect the photospheric
emission to deviate from a simple Planck function, and several authors
have recently considered models that give rise to photospheric
emission with a broad, Band-like spectrum
(\citealt{Peer06,Beloborodov09,Lazzati10,Mizuta10}). A broadened
photospheric component is also to be expected from the simple fact
that the observed spectrum is most likely composed of emission arising
from different regions in space. As a result, the angle dependence of
the optical depth and the Doppler shift, as well as possible
angle-dependent density and Lorentz-factor profiles, may all
contribute to create the broadening (\citealt{Peer07,Peer10}).

Here we analyse the bright, long burst \1007 with the aim of exploring
models in which the emission is dominated by a photospheric
component. The prompt emission of \1007 was caught by both \swift and
\suzaku (\citealt{Ohno09}) as well as {\it Konus-Wind}
(\citealt{Golenetskii06}). With an isotropic energy release of
\hbox{$\sim10^{54}$~erg} in the \hbox{1-10~000~keV} energy band
(\citealt{Ohno09}) this burst is among the most energetic GRBs ever
observed, comparable to the brightest bursts detected by the \fermi
LAT (e.g.~\citealt{Cenko10}). A very bright afterglow was observed by
\swift and several ground-based telescopes
(\citealt{Mundell07,Schady07}), and the redshift was measured to be $z
= 1.261$ (\citealt{Osip06}, \citealt{Jakobsson06}).

Due to its brightness and the wide energy band offered by the combined
\swift and \suzaku observations, \1007 is an ideal candidate for
detailed time-resolved studies of the gamma-ray emission. The \suzaku
and \swift data were previously analysed by \cite{Ohno09}, who carried
out a time-resolved spectral analysis on a 1-s time-scale. Modelling
the spectra with a Band function, the authors found that the
time-resolved spectra follow the same $E_{\rm{peak}} - L_{\rm{iso}}$
relation as the time-averaged spectra of other bursts
(\citealt{Yonetoku04}), but also noted that the initial rising phases
of the pulses may be outliers to this relation.

Here we find that that an alternative interpretation in terms of
photospheric emission is possible, and explore the consequences of
this in terms of the evolution of the fireball properties. In addition
to a thermal component, our best-fitting spectral model also requires
the presence of a power law. We discuss the properties of this
component in view of recent \fermi results. This paper is organized as
follows.  We describe the observations in section \ref{observations}
and present the time-resolved analysis in section \ref{analysis}. We
then explore the properties of the relativistic outflow as derived
from our spectral fits in section \ref{photosphere}.  A discussion and
our conclusions are presented in section \ref{discussion}.

\section{Observations and data reduction}
\label{observations}

\1007 was observed with the \suzaku WAM and the \swift BAT
detectors. The $T_{90}$ duration was measured to be 59~s in the
50--5000~keV WAM energy range (\citealt{Yamaoka06}) and 75~s in the
150--500~keV BAT energy range (\citealt{Schady07}). The observations
are described in detail in \cite{Ohno09}, and for this work we also
follow the data reduction procedure described in that paper. For the
WAM analysis we thus use the WAM-3 detector, in which the burst was
most strongly detected. For both the WAM and the BAT we extract
science products on a 1~s time-scale, as this corresponds to the
temporal resolution of the transient class WAM data.

\begin{figure}
\begin{center}
\rotatebox{270}{\resizebox{!}{80mm}{\includegraphics{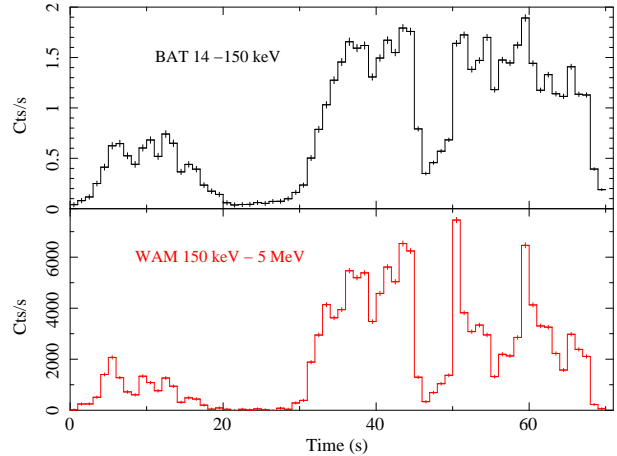}}}
\caption{\small{BAT + WAM light curves on a 1 s time-scale. The BAT
    count rates are per fully illuminated detector for an equivalent
    on-axis source.}}
\label{lc}
\end{center}
\end{figure}

\section{Time-resolved spectral analysis}
\label{analysis}

Fig.~\ref{lc} shows the WAM and BAT light curves of \1007 on a 1~s
time-scale. The light curve covers 70 seconds and three main pulses can
clearly be identified in both energy bands. The time bins shown in the
light curve are also used for the spectral analysis, excluding the
first and last bins as well as the time between 19 and 28 seconds, due
to the low count rates during these times. This leaves us with a total
of 58 spectra with 1-s duration for the analysis.

The spectra were fitted over the 14--150~keV (BAT) and 150--5000~keV
(WAM) energy ranges, with the cross-normalization between the two
detectors left as a free parameter (as in \citealt{Ohno09}).  Fixing
the cross-normalization at the value obtained from the time-averaged
spectra only changes the results marginally, as discussed in section
\ref{crossnorm}. All fits were performed using {\scriptsize XSPEC}12,
assuming a standard cosmology with $\Omega_{\lambda} =
0.73,\ \Omega_{\rm{M}} = 0.27$ and $H_{0} =
71\ \rm{km\ s^{-1}\ Mpc^{-1}} $. Errors on model parameters and error
bars in plots represent the one sigma confidence level for one
interesting parameter, unless otherwise stated.

\subsection{Identification of a photospheric component}

As a first step we investigate whether the spectra can be fitted with
a model in which the peak of the emission is provided by a single
Planck function. To account for any non-thermal emission we also
include a second component in our models. We consider two forms for
this component, a power law and cutoff power law. For the majority of
the spectra the blackbody+power-law model does not provide an
acceptable fit. Replacing the power law with a cutoff power law
results in excellent fits, but we note that in this case the
contribution from the blackbody is very small. This still holds
if the energy of the blackbody is constrained in order to correspond
to the peak of the spectra.  We thus conclude that the prompt emission
of this burst is not dominated by a single blackbody component.

As mentioned in section \ref{introduction}, another, more realistic,
possibility is that the photospheric emission is broadened. We model
this scenario with a multicolour blackbody, using the {\scriptsize
  XSPEC} model \textit {diskpbb}. This model was developed to describe
emission from an accretion disc, but the parameters can also be
interpreted in the general case of multicolour blackbody emission. The
free parameters of the model are the shape of the blackbody (described
by the parameter $p$), the maximum blackbody temperature ($T_{\rm{max}}$)
and the normalization. More specifically, the parameter $p$ describes
the temperature profile of an accretion disc giving rise to
multicolour blackbody emission. Since this is irrelevant in the case
of GRBs, we instead introduce the parameter $q$, which relates to the
fitted parameter $p$ as $q = 4 -2/p$. Using $q$ we can express the
relationship between the flux and the temperature of the single
blackbody components (which together make up the multicolour
blackbody) as
\begin{equation}
 F(T) = F_{\mathrm {max}} \left(\frac{T}{T_{\mathrm {max}}}\right)^q ,
\label{bbfluxes}
\end{equation}
where $F_{\mathrm {max}}$ is the flux of the Planck function at
$T_\mathrm {{max}}$.  As $q \rightarrow \infty$ we approach the case
of a single blackbody. We note that the same parameter $q$ is used to
describe the multicolour-blackbody model in \cite{Ryde10}. We also use
(\ref{bbfluxes}) to define the average temperature of the multicolour
blackbody
\begin{equation}
T = \frac{\int_0^{T_\mathrm {{max}}} T  F(T) dT}{\int_0^{T_\mathrm
    {{max}}} F(T) dT} = T_\mathrm {{max}}  \frac{q+1}{q+2} \ .
\end{equation}

\begin{figure}
\begin{center}
\resizebox{!}{80mm}{\includegraphics{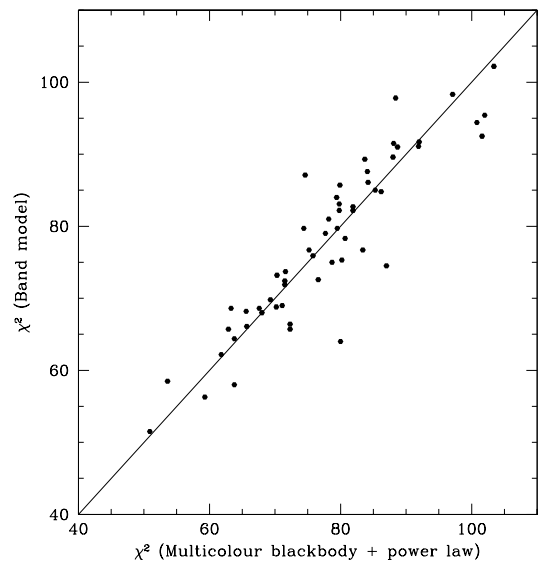}}
\caption{\small {Comparison of $\chi^2$ for the Band and
    multicolour-blackbody models. Both models have five free
    parameters.}}
\label{cfband}
\end{center}
\end{figure}

\begin{figure}
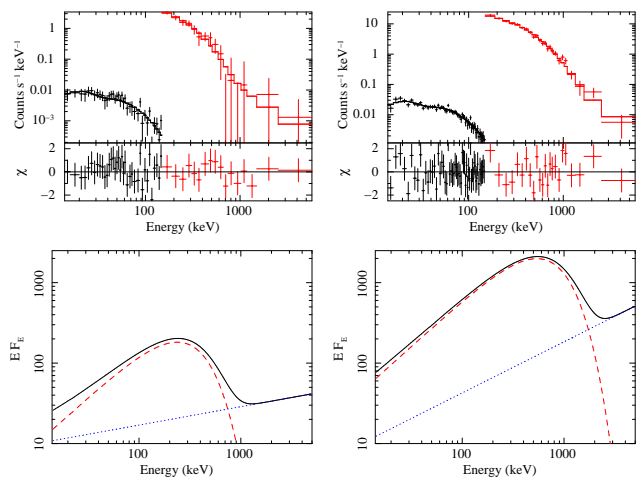

\begin{center}
\rotatebox{270}{\resizebox{!}{40mm}{\includegraphics{fig3a.eps}}}
\hspace{0.1cm}
\rotatebox{270}{\resizebox{!}{40mm}{\includegraphics{fig3b.eps}}}\\
{\vspace{0.18cm}}
\rotatebox{270}{\resizebox{!}{40mm}{\includegraphics{fig3c.eps}}}
\hspace{0.1cm}
\rotatebox{270}{\resizebox{!}{40mm}{\includegraphics{fig3d.eps}}}
\caption{\small{Sample fits to the mBB+pl model. Plots in the left
    column are for the spectrum at time 16~s (at low flux) and the
    right column are for the spectrum at time 41~s (at high flux). The
    top row shows the count spectra and the residuals of the fit, with
    BAT data shown in black and WAM data shown in red. The bottom row
    shows the deconvolved components of the best-fit models. The
    solid, black line is the total model, the dashed, red line is the
    multicolour blackbody and the dotted, blue line is the power
    law.}}
\label{spectra}
\end{center}
\end{figure}

To allow for an additional non-thermal component we also include a
power law in the model, $ N(E) = K E^{s}$, where $K$ is the
normalization at 1~keV in photons~$\rm{keV^{-1}\ cm^{-2}\ s^{-1}}$ and
$s$ is the photon index. The resulting multicolour-blackbody +
power-law (mBB+pl) model provides an excellent fit to all the
spectra. Interestingly, we also find that the shape of the multicolour
blackbody stays remarkably constant, $p_{\rm{av}}=0.70 \pm0.06$
(corresponding to $q=1.1$), based on the 47 spectra for which no
problems occurred in the error calculations. We therefore re-fit the
spectra with $p$ fixed at $0.70$ in order to obtain better constraints
on the remaining model parameters. Fig.~\ref{cfband} shows the
resulting values of $\chi^2$ compared to those obtained from fits to a
Band model (both models have 80 degrees of freedom). Clearly it is not
possible to distinguish between the two models on a statistical
basis. The cross-normalization ($C$) between the WAM and the BAT is also
similar for the two models ($C =1.27 \pm 0.26$ for Band and $C =1.17\pm
0.19$ for mBB+pl), showing that this parameter does not significantly
affect the results (see also the discussion in section
\ref{crossnorm}). Examples of fits to the mBB+pl model are shown in
Fig.~\ref{spectra}.

\begin{figure*}
\begin{center}
\resizebox{!}{80mm}{\includegraphics{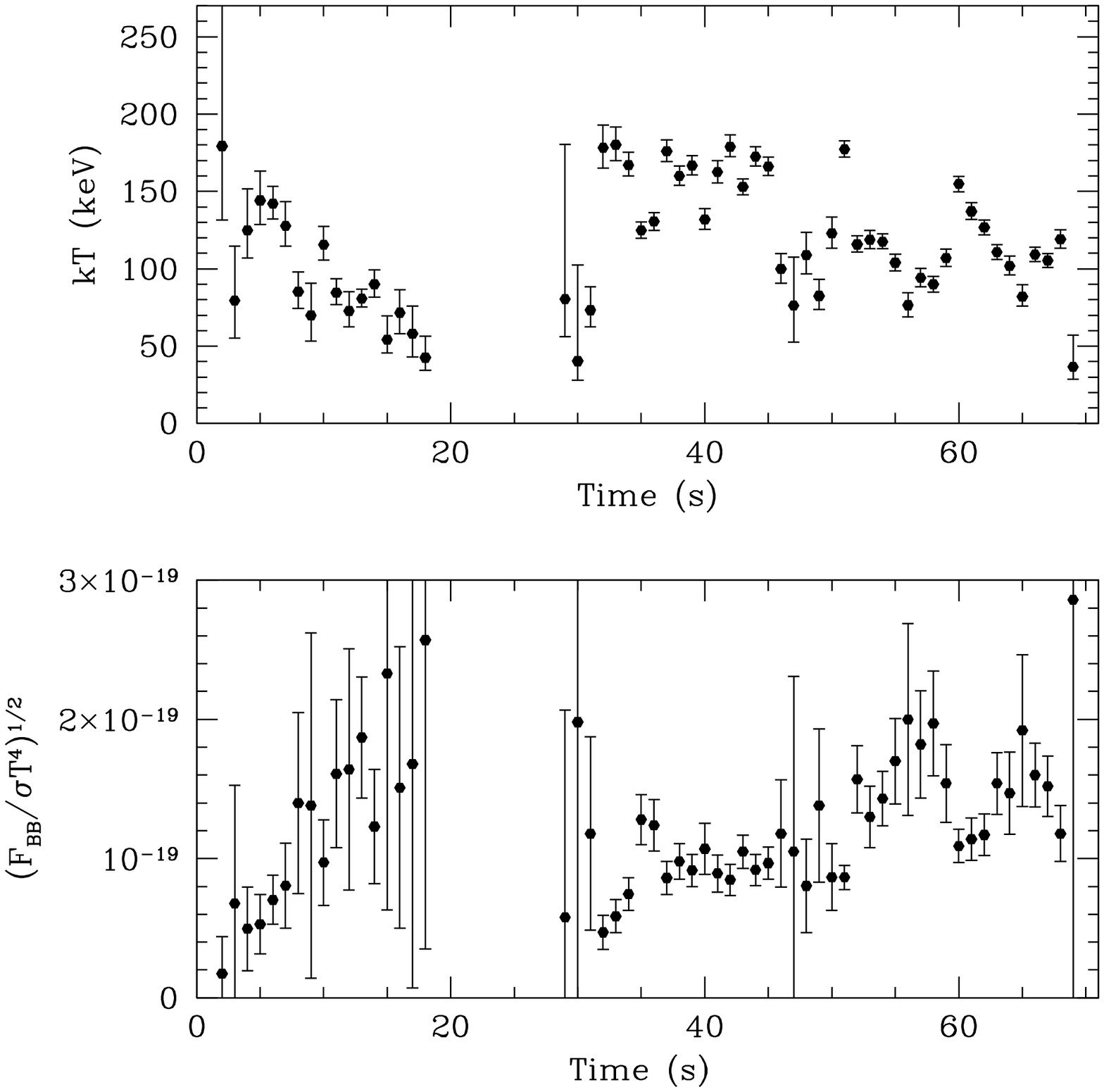}}
\resizebox{!}{80mm}{\includegraphics{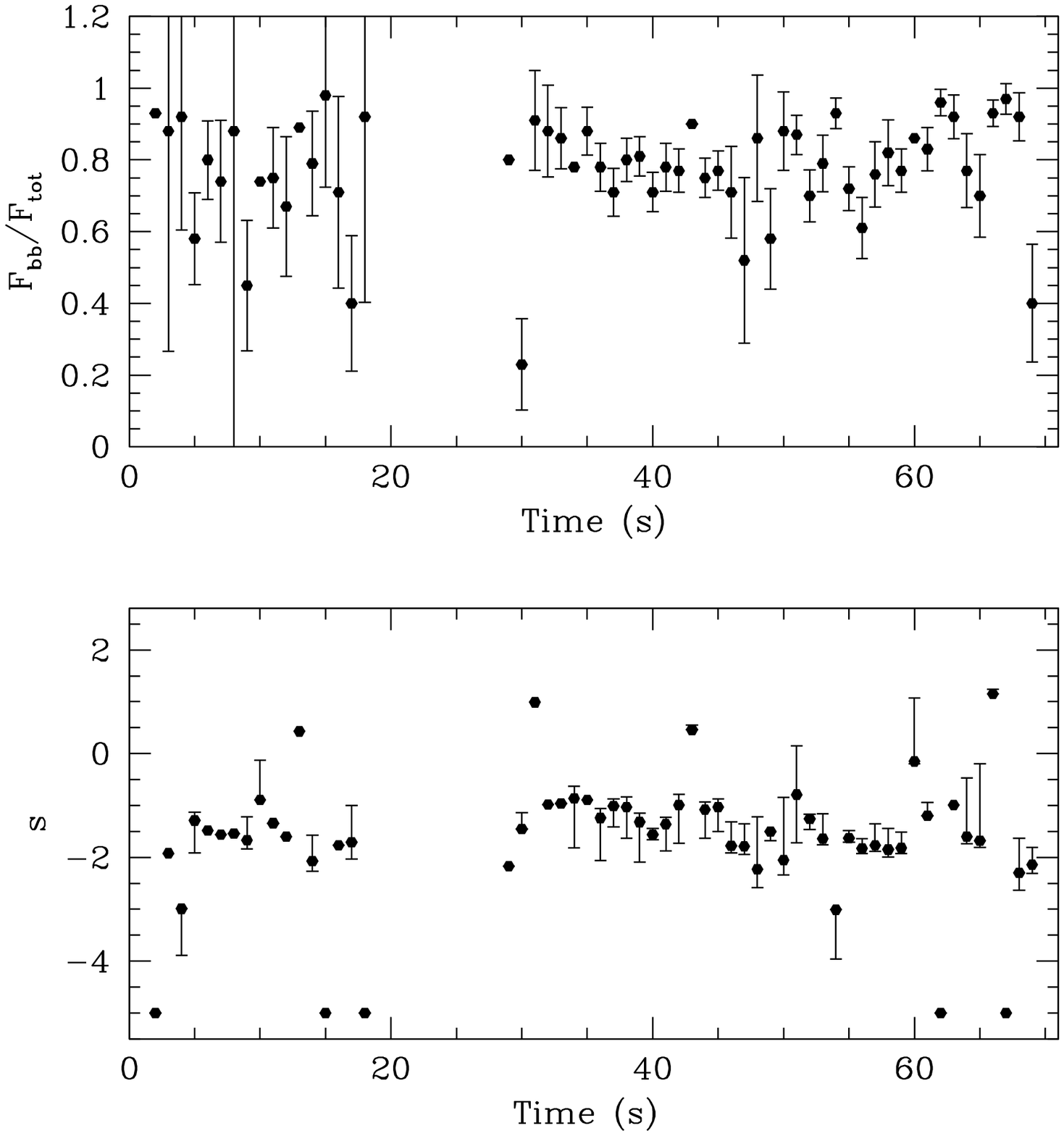}}
\caption{\small{Results from the mBB+pl fits as a function of
    time. The shape of the blackbody was kept fixed at $p=0.70$. Top
    left: evolution of the temperature of the multicolour
    blackbody. Bottom left: the effective size of the emission
    region. Top right: contribution of the blackbody to the total
    flux. Bottom right: power-law photon index. Data points without
    error bars indicate that the error calculation did not converge.}}
\label{mbbresults}
\end{center}
\end{figure*}

The evolution of $kT$ with time is shown in the top left panel of
Fig.~\ref{mbbresults}. A comparison with the light curve in
Fig.~\ref{lc} reveals that $kT$ correlates with the flux. We will
explore this relation in more detail in section \ref{pulses} below. In
the lower left panel of the same figure we plot the dimensionless
quantity $(F_{BB} / \sigma T^4)^{1/2}$, which is a measure of the
effective size of the emitting region. This parameter is seen to
increase during the rising phase of each pulse and then only vary
mildly during the decaying phase. In particular, the variations in
this parameter are much smaller than the variations of the light curve
(Fig~\ref{lc}), supporting the interpretation that this burst is
dominated by a photospheric component.

In Fig.~\ref{mbbresults} we also plot the ratio of the blackbody flux
to the total flux and the power-law photon index. We see that the
contribution from the blackbody component is fairly stable at around
75 per cent of the flux, although less well constrained during the
first pulse. Due to the low flux of the power-law component in the
fitted energy interval, the photon index could not be constrained in
all the spectra. Considering only those 31 spectra for which it could
be constrained, we find that a fit to a constant gives
$s_{\rm{av}}=-1.50$ with $\chi^2 / d.o.f. = 30.8/30$.
  
\subsection{Significance and properties of the power law}

It is interesting to consider the power-law component of our model in
view of recent \fermi observations, which have revealed additional
power-law components at high energies in a number of bright GRBs
(e.g.~\citealt{Abdo09,Ackermann10}). We especially note that our
photon index values are clustered around $-1.5$, which has also been
observed in several \fermi bursts (\citealt{Guiriec10}) and which is
indicative of synchrotron or inverse Compton emission, as discussed in
Section \ref{discussion}. However, we have also noted that the
contribution from the power law to the total model is small and that
the photon index cannot be constrained in many of the time intervals,
which prompts the question of whether the power law is required in the
fits.

\begin{figure}
\begin{center}
\rotatebox{270}{\resizebox{!}{80mm}{\includegraphics{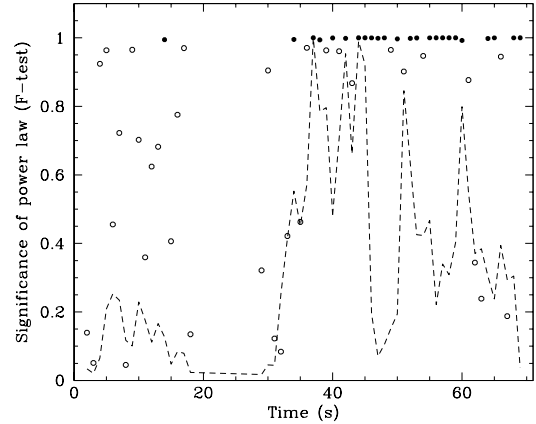}}}
\caption{\small{Significance of the power law according to an
    F-test. Filled circles indicate spectra for which the power law is
    required above 99 per cent confidence. The dashed line is the
    light curve of the burst.}}
\label{ftest}
\end{center}
\end{figure}

To test this we fit all spectra with a pure multicolour-blackbody
model ($p$ still fixed at 0.70) and use an F-test to compare these
fits with the mBB+pl ones. The results are plotted in
Fig.~\ref{ftest}. We see that the significance of the power law is
well above 90 per cent in the majority of the spectra, but that the
significance is lower at low fluxes. As expected, the photon index
is also poorly constrained in these low-flux intervals
(cf.~Fig.~\ref{mbbresults}). In order to make sure that the high
significance of the power law in most of the spectra is not due to the
fact that $p$ was kept fixed, we also carry out the same test after
fitting the spectra with a multicolour blackbody with $p$ as a free
parameter. In this case the overall significance of the power law is
somewhat weaker, but it is still required at very high significance in
the majority of spectra after the initial, weak pulse.

\begin{figure}
\begin{center}
\resizebox{80mm}{!}{\includegraphics{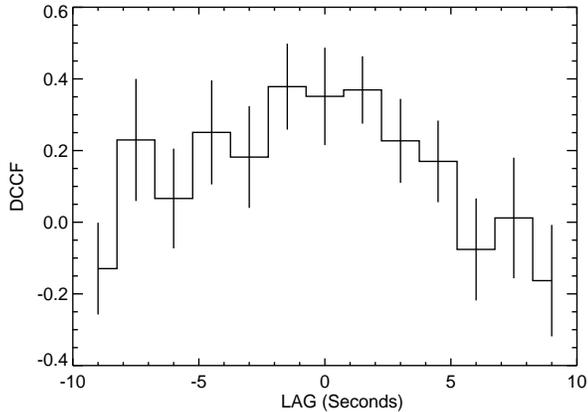}}
\caption{\small{The cross correlation function between the blackbody
    and power-law components. Negative lag means that the power law
    flux leads that of the blackbody.}}
\label{tcorr}
\end{center}
\end{figure}

Based on our spectral fits we also derive flux light curves for both
the multicolour blackbody and the power-law component. The correlation
between these light curves was investigated by calculating the discrete
cross correlation function (DCCF) following \citet{Edelson88}. The
resulting DCCF is shown in Fig~\ref{tcorr}. The relative lag and the
correlation strength were computed by fitting a Gaussian profile to
the correlation peak in the DCCF, and uncertainties in these
parameters were estimated using a Monte Carlo method (see
\citealt{Peterson98}). The result of this analysis is a correlation
strength of 0.44 +/-0.14 (corrected for the effect of measurement
noise) and a lag which is consistent with zero, -0.8 +/-1.5 sec, where
negative lag means that the power law leads the blackbody component.

\subsection{Spectral evolution during individual pulses}
\label{pulses}

\citet{Ohno09} has already considered the $E_{\rm{peak}} -
L_{\rm{iso}}$ evolution of this burst as derived from fits to a Band
function. They found that the time-resolved spectra satisfy the
$E_{\rm{peak}} - L_{\rm{iso}}$ relation defined by time-averaged
bursts (\citealt{Yonetoku04}), but that a number of points in the
initial rising phases of the three pulses are outliers to this
relation. Here we instead consider the spectral evolution in terms of
the flux and temperature of the multicolour blackbody, which provides
a natural explanation for the relation between peak energy and
luminosity.

\begin{figure*}
\begin{center}
\resizebox{70mm}{!}{\includegraphics{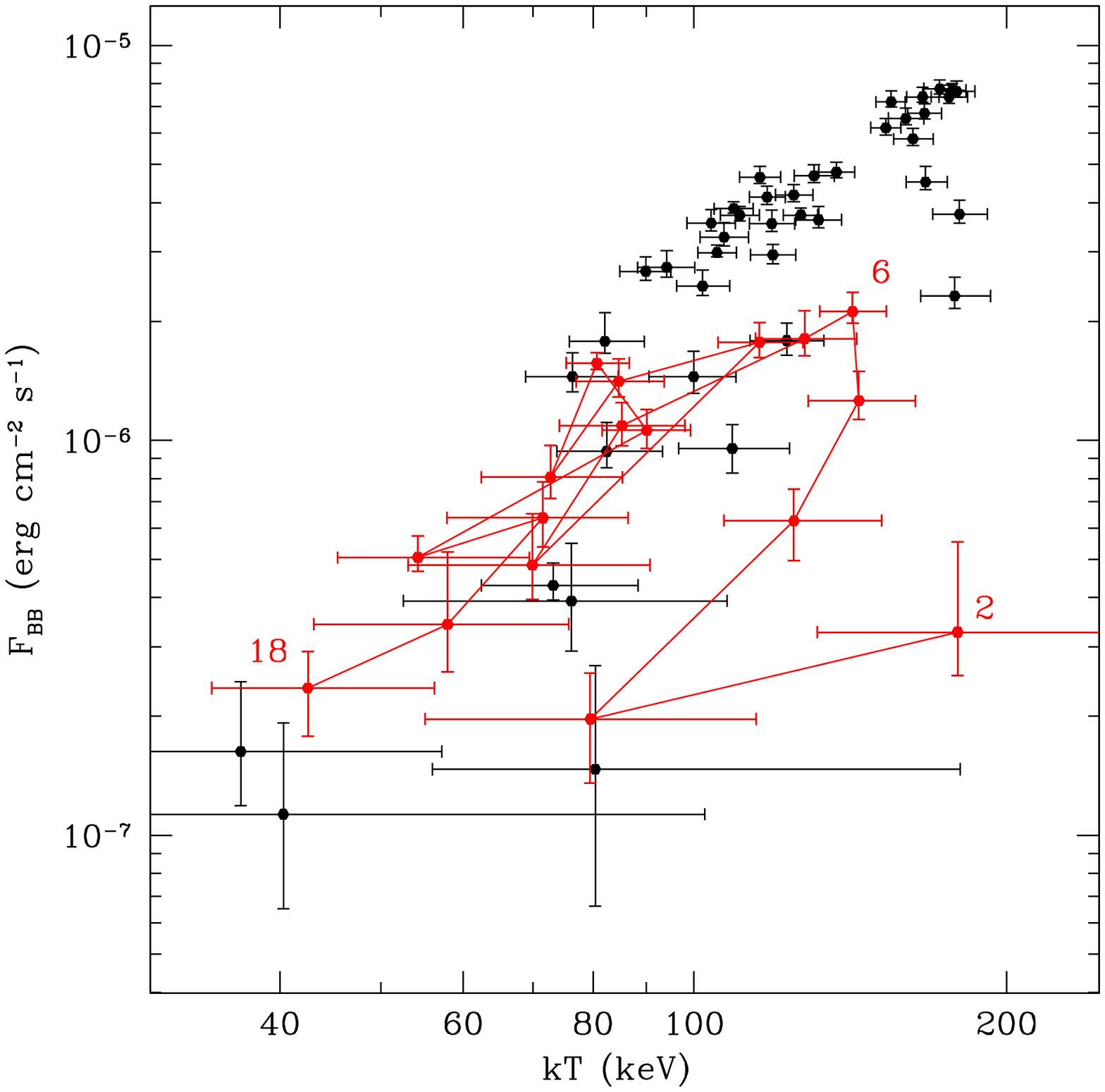}}
\resizebox{70mm}{!}{\includegraphics{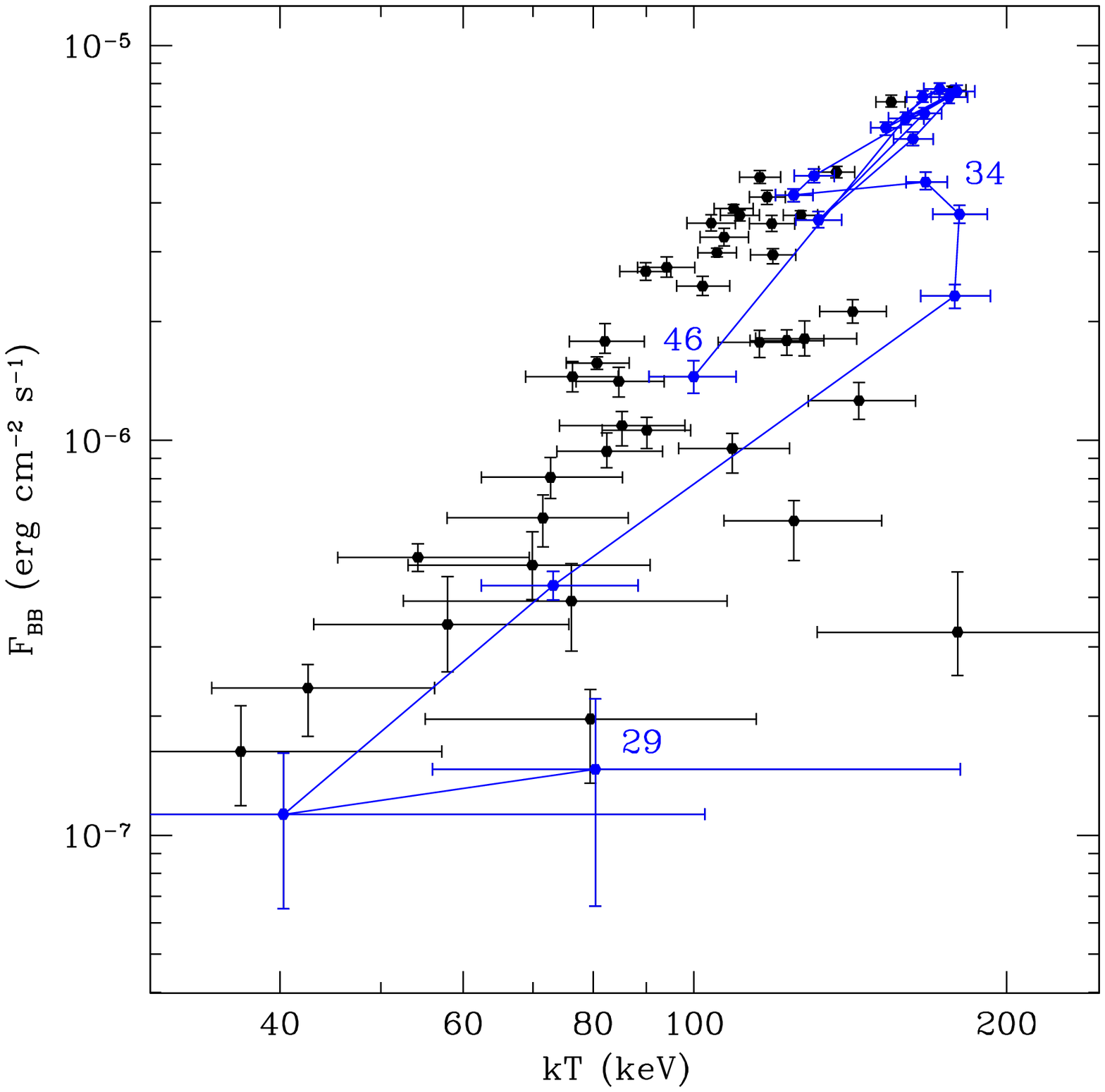}}
\resizebox{70mm}{!}{\includegraphics{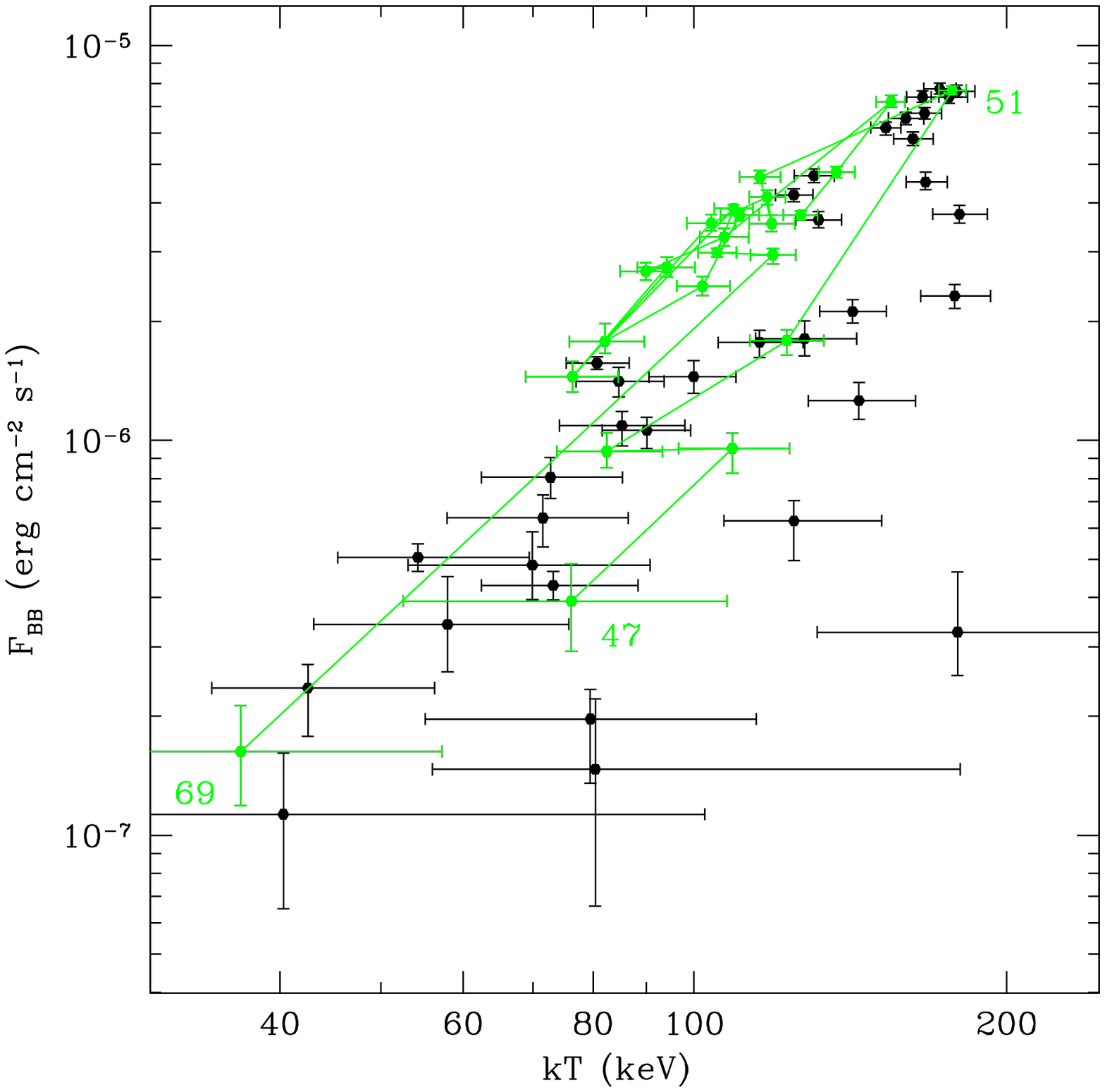}}
\resizebox{70mm}{!}{\includegraphics{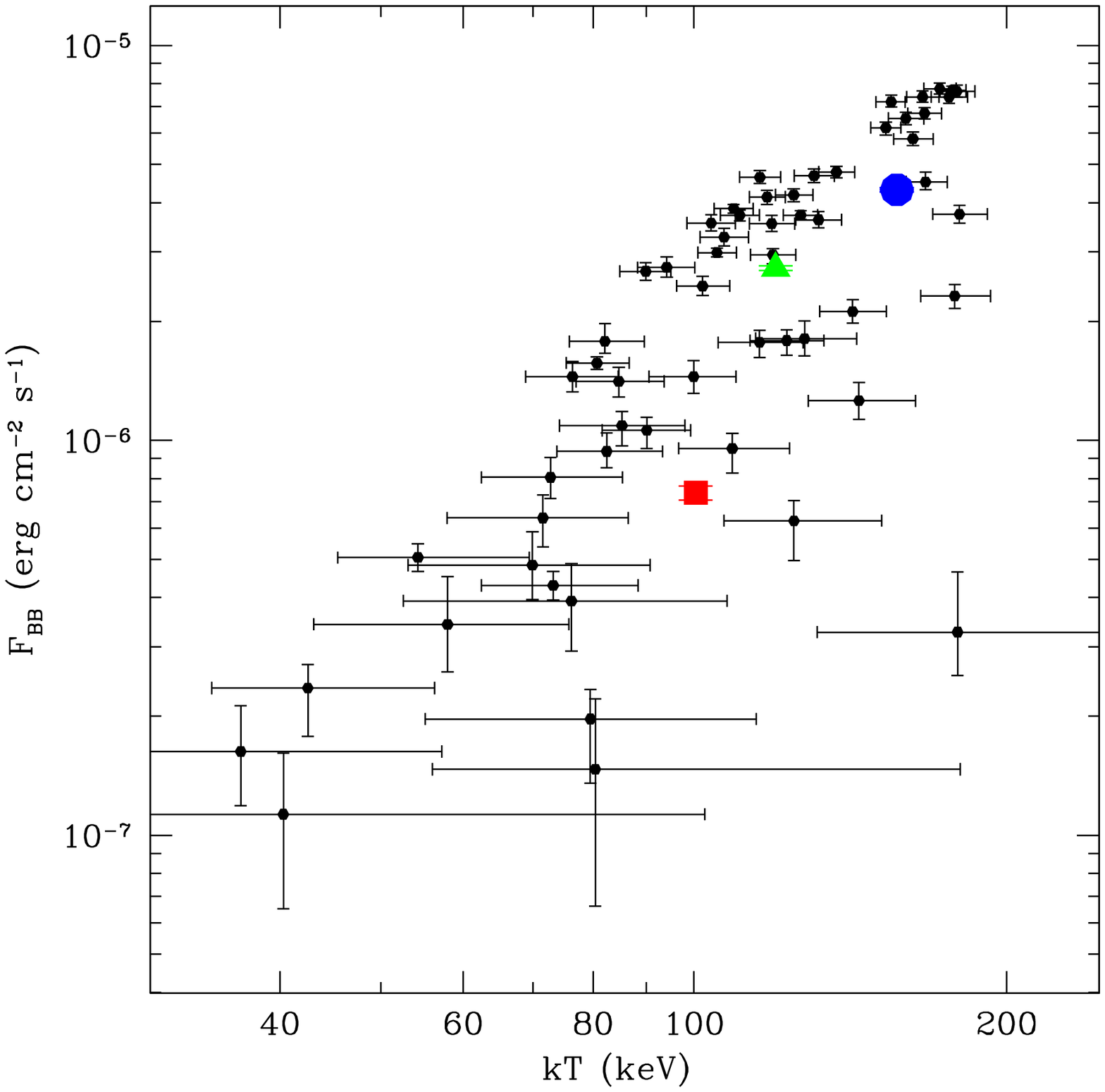}}
\rotatebox{270}{\resizebox{!}{130mm}{\includegraphics{fig7e.eps}}}
\caption{\small{Evolution of the flux and temperature of the
    multicolour blackbody during individual pulses. The first pulse is
    tracked in red in the top left panel, the second pulse is tracked
    in blue in the top right panel and the third pulse is tracked in
    green in the middle left panel. The middle right panel shows the
    results of time-averaged fits to the three pulses as a red square,
    a blue filled circle and a green triangle, respectively. The
    bottom panel shows the light curve with the three pulses
    identified. The initial rising phase of each pulse clearly follows
    a lower track than the rest of the pulse. In addition, we see that
    each pulse starts rising on a slightly higher track than the
    previous one.}}
\label{fkt}
\end{center}
\end{figure*}

The temporal evolution of each pulse in the $F_{\rm{BB}} - kT$
representation is shown in Fig.~\ref{fkt}. The pulses exhibit a
tracking behaviour, with the rising phase of each pulse moving along a
lower track in the diagram. In addition, each pulse starts rising
slightly higher up than the previous one. To quantify the temporal
evolution within the pulses we fit straight lines to the logaritmized
data in the rising and decaying phases of each pulse, using a fitting
routine that accounts for the errors in both coordinates. We find that
the correlation indices of the decaying phases are clustered around
2.2-2.3, which is typical for decaying phases of GRBs
(e.g.~\citealt{Borgonovo01}, \citealt{Lu10}). The rising phases have
steeper indices (values around 3) but are poorly constrained due to
the small number of data points.

As the three main pulses are clearly separated we can also carry out
the experiment of treating them as individual bursts, and fit their
time-averaged spectra with the mBB+pl model. The results of the three
time-averaged fits fall within the range of results from the 1-s fits,
as shown in Fig.~\ref{fkt}. This is to be expected in the picture
where the the various hardness-intensity correlations between the
time-averaged spectra of different bursts are driven by the behaviour
that is seen within individual bursts (see e.g~\citealt{Firmani09}).

\subsection{Impact of the cross-normalization between the WAM and the BAT}
\label{crossnorm}

In the time-resolved fits described above we have allowed the
cross-normalization factor ($C$) between the WAM and the BAT to be a
free parameter. $C$ was also left free in the previous analysis of this
burst carried out by \cite{Ohno09}. Since the \suzaku satellite kept
the same attitude during the burst the $C$-values obtained from the
time-resolved fits should in principle be consistent with the $C$-value
from a time-averaged fit. However, we find that fixing $C$ at the
time-averaged value results in significantly worse $\chi^2$ in the
intervals with high fluxes, regardless of the type of model being
fitted. This is most likely due to systematic errors in the WAM
detector, which increase with the brightness of the burst.

As an alternative to keeping $C$ free we therefore performed all the
time-resolved fits with $C$ fixed at the time-averaged value, but with 3
per cent systematic errors added to the WAM energy channels. The
quality of the resulting fits is very similar to the fits with $C$
free. The only systematic change in terms of best-fitting parameters
is that $p$ is slightly higher, $p_{\mathrm{av}}=0.73 \pm 0.08$
compared to $p_{\mathrm{av}}=0.70 \pm 0.06$ for $C$ free, which
translates into a slightly narrower multicolour blackbody.

\section{Properties of the fireball}
\label{photosphere}

\begin{figure}
\begin{center}
\rotatebox{270}{\resizebox{!}{80mm}{\includegraphics{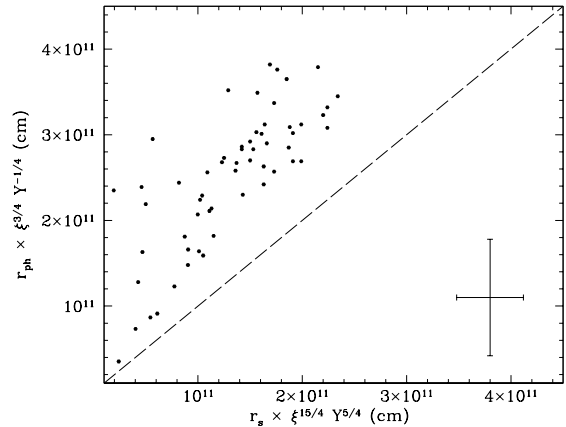}}}
\caption{\small{The photospheric radius versus the saturation radius
    of the fireball. Typical one-sigma error bars are shown in the
    lower right corner.}}
\label{rphot}
\end{center}
\end{figure}

\begin{figure*}
\begin{center}
\resizebox{!}{80mm}{\includegraphics{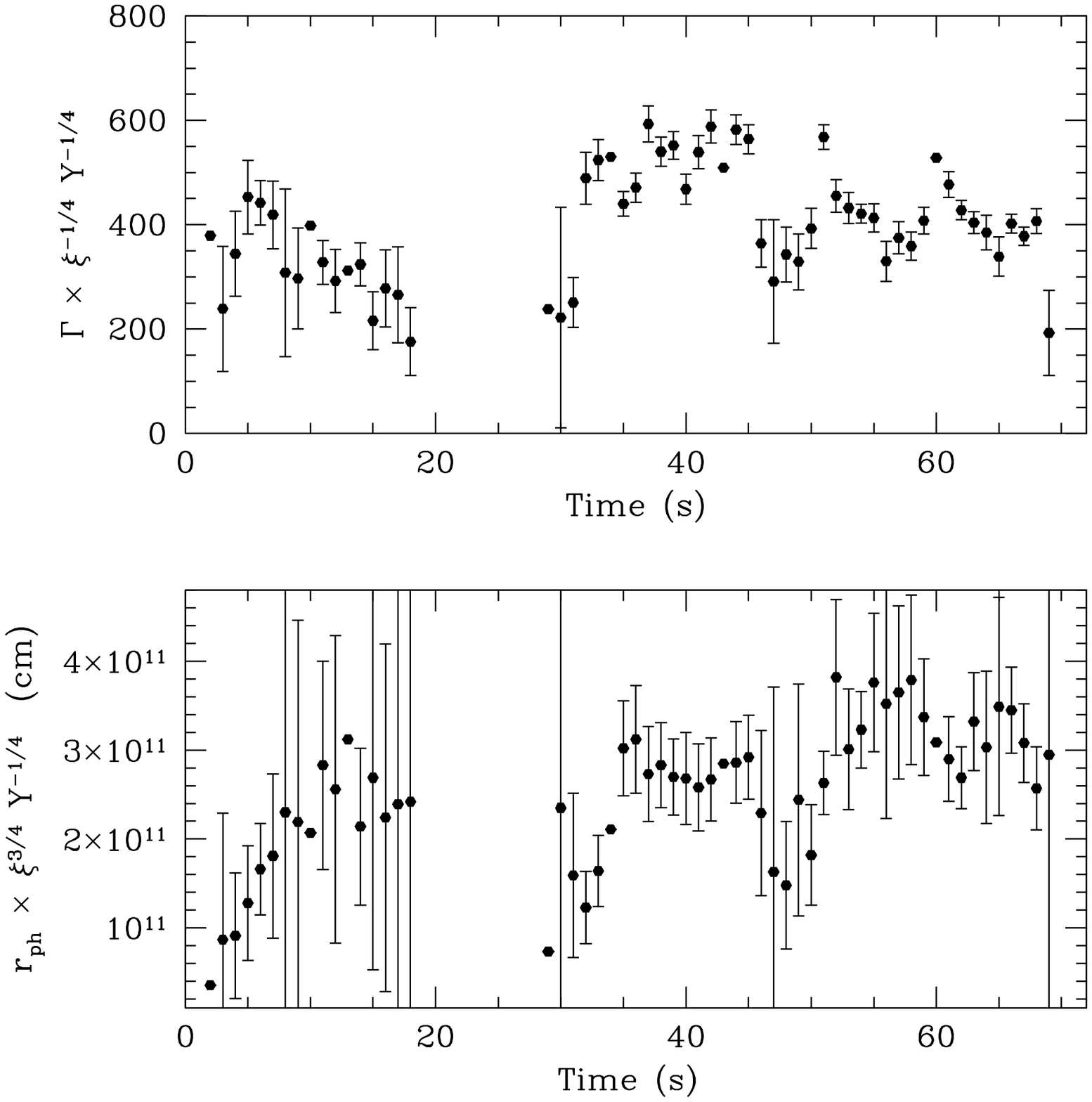}}
\resizebox{!}{80mm}{\includegraphics{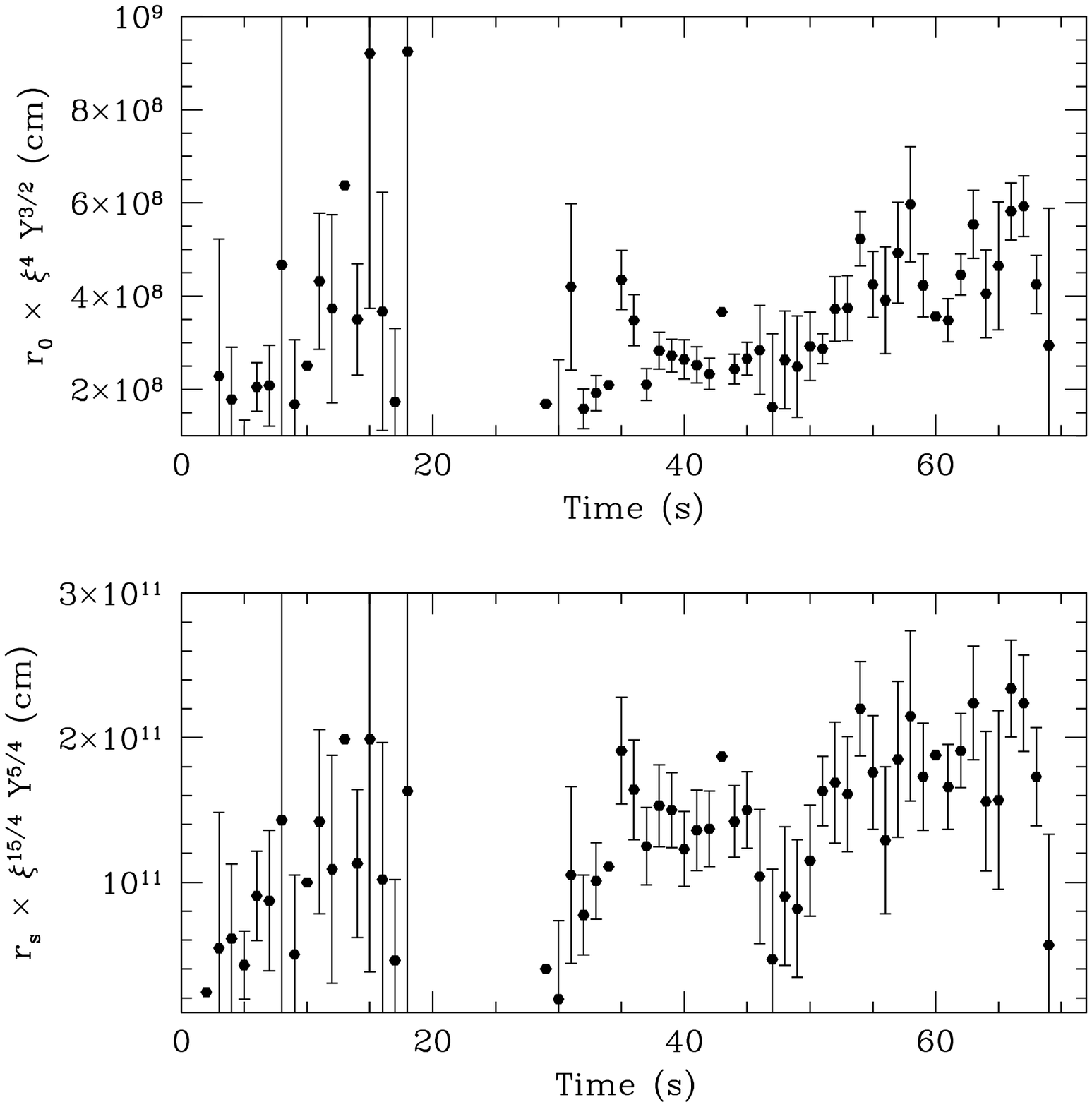}}
\caption{\small{Properties of the outflow, calculated from the
    best-fitting parameters of the mBB+pl model. $\xi$ is a
    geometrical factor of order unity and $Y$ is the ratio between the
    total fireball energy and the energy emitted in gamma
    rays. $\Gamma$ is the bulk Lorenz factor, $r_{\rm{0}}$ is the size at
    the base of the flow, $r_{\rm{ph}}$ is the radius of the photosphere
    and $r_{\rm{s}}$ is the saturation radius.}}
\label{fireball}
\end{center}
\end{figure*}

Having identified a photospheric component in the spectra, we can now
use the measured temperature and flux of this component to determine
the properties of the relativistic outflow. Assuming the
non-dissipative fireball model described in \cite{Peer07} we calculate
the bulk Lorentz factor of the flow ($\Gamma$), the radius of the
photosphere ($r_{\rm{ph}}$), the size at the base of the flow ($r_{\rm{0}}$) and
the saturation radius ($r_{\rm{s}}$). In order to determine all these
parameters we must have $r_{\rm{ph}} > r_{\rm{s}}$, which we confirm following
the procedure in \cite{Peer07} (see also
Fig.~\ref{rphot}). Fig.~\ref{fireball} shows all the parameters of the
outflow as a function of time.

$\Gamma$ is seen to vary between about 200 and $600~(\xi Y)^{1/4}$,
where $\xi$ is a geometrical factor of order unity and $Y$ is the
ratio between the total fireball energy and the energy emitted in
gamma rays. The highest values of $\Gamma$ are seen during the second
pulse. The other three parameters plotted in Fig.~\ref{fireball}
($r_{\rm{ph}}$, $r_{\rm{0}}$ and $r_{\rm{s}}$) all show a trend of increasing during
the rising phase of each pulse and then remaining nearly constant
during the decaying phase. The average values of these parameters are
$r_{\rm{ph}} = 2.5 \pm 0.8 \times
10^{11}\ \xi^{-3/4}\ Y^{1/4}\ \mathrm{cm}$, $r_{\rm{0}} = 3.5 \pm 1.7
\times 10^{8}\ \xi^{-4}\ Y^{-3/2}\ \mathrm{cm}$ and $r_{\rm{s}} = 1.3 \pm
0.6 \times 10^{11}\ \xi^{-15/4}\ Y^{-5/4}\ \mathrm{cm}$. It should be
noted that there is some uncertainty in the last two parameters as
they depend on the assumed fireball dynamics during the acceleration
phase (see \citealt{Peer07} for details).

There is also a systematic uncertainty in our estimates of all these
parameters due to the fact that they all depend on the total gamma-ray
flux. For the results presented in Fig.~\ref{fireball}, we have simply
used the flux in the observed BAT+WAM energy band, but the flux will
of course be higher if the power-law component extends to high
energies. The effect of a higher total gamma-ray flux would be higher
values of $\Gamma$ and $r_{\rm{ph}}$ and lower values of $r_{\rm{0}}$
and $r_{\rm{s}}$, with the change being significantly larger for the
latter two parameters.

In order to estimate to what energy the power law is likely to extend
we consider the value of $r_{\rm{0}}$. As this is the initial size of
the fireball we would expect a value of about $10^7$~cm, i.e.~the
Schwarzschild radius of a solar-mass black hole. We therefore
determine where the power laws from all our spectral fits would need
to break in order to be consistent with this value. Assuming a break
to a slope of $-2$ we find that the majority of the spectra should
have a break in the 10~MeV - 1~GeV band, while a sharp cutoff would
give breaks in the 100~MeV - 100~GeV range. If we also use the
inferred power-law breaks to calculate new values of the other
parameters, we find that $\Gamma$ and $r_{\rm{ph}}$ are up to a factor
of two higher than in Fig.~\ref{fireball}, while $r_{\rm{s}}$ is
between 10 and 50 times smaller.

\section{Discussion and conclusions}
\label{discussion}

After the launch of \fermi it has become increasingly apparent that
GRB spectra are composed of multiple components. While some spectra
are still very well described by the canonical Band function, others
require an additional power-law component extending to high energies
and/or a photospheric component (e.g. \citealt{Zhang10}).  The power
law has been detected with high significance in a handful of bursts
(\citealt{Abdo09,Ackermann10,Guiriec10,Ackermann11}) but there is
marginal evidence in many more (\citealt{Granot10}). A photospheric
component is clearly seen in both GRB~090902B (together with a
power-law component, \citealt{Ryde10}) and GRB~100724B (together with
a Band function, \citealt{Guiriec10b}). In the case of GRB~090902B the
photospheric component is best modelled by a multicolour
blackbody. Given this evidence it is natural to ask whether such
composite models can explain the prompt spectra observed with other
satellites than \fermi.  We have found that this is indeed the case
for the \swift + \suzaku data of GRB 061007, which can be equally well
fitted by a multicolour blackbody together with a power law as by a
Band function. Below we discuss the properties and possible origin of
the photospheric and power-law components.

\subsection{The photospheric component}

We find that the multicolour blackbody dominates the prompt emission
of GRB~061007, providing about 75 per cent of the total flux
throughout the burst. Since the luminosity and temperature of a
blackbody are related this model also provides a natural explanation
for the hardness-intensity correlation within the burst. This is a
major advantage compared to models where the bulk of the emission is
interpreted as synchrotron emission, where these correlations are much
harder to explain (e.g. \citealt{Ramirez-Ruiz02}).

Another interesting result from our fits is that the shape of the
multicolour blackbody is consistent with staying constant throughout
the burst. The same shape is also found when fitting the time-averaged
spectra of the three main pulses. This result disfavours a scenario
where the broadening is primarily due to rapid temperature
fluctuations, as this should give rise to a narrowing of the
multicolour blackbody when shorter time intervals are considered (as
in GRB 090902B, \citealt{Ryde10,Zhang10}).

Several other mechanisms that may create broadened photospheric
spectra have recently been discussed in the literature. These include
sub-photospheric heating due to collisions between neutrons and
protons (\citealt{Beloborodov09}) or shocks or magnetic dissipation
(\citealt{Lazzati10}). These models result in high-energy slopes that
match those of typical Band spectra, but fail to explain the observed
range of low-energy slopes. In order to explain the low-energy part of
the multicolour blackbody in GRB 061007 (which can be approximated by
a power law with photon index $\alpha \approx -0.8$) a different
explanation is clearly needed. The most important effects that could
contribute to a softer low-energy slope can be summarized as follows:

\begin{enumerate}
\item The observed temperature depends on the Doppler boost, which in
  turn depends on the angle to the line-of-sight ($T_{\rm{ob}} = D T' = T'
  / \Gamma (1-\beta cos\theta)$, where $T'$ is the outflow temperature
  measured in the comoving frame). Integrating the emitting surface
  over angle will therefore produce a multicolour blackbody
  spectrum. \cite{Peer10} have shown that for late times, high
  latitude effects produce a spectrum with $\alpha = -1$.
\item The assumption that all photons originate from $r(\tau=1)$ is
  not always a good approximation (\citealt{Peer08}). Depending on how
  the comoving outflow-density scales with radius, the probability
  density function for a photon to make its last scattering at radius
  $r$ does not have to be sharply peaked around $r(\tau = 1)$. If the
  local outflow temperature or Lorentz factor varies on distance
  scales similar to the width of the probability-density profile, an
  integration over radius is necessary to obtain the observed
  spectrum.
\item The outflow properties around the photosphere are expected to
  vary on a time-scale shorter than the most highly time-resolved
  spectra (\citealt{Rees05}). Thus, observed spectra are likely the
  result of an integration over time, which can produce a soft
  low-energy slope for reasonable temporal scalings of the emitting
  surface and the temperature (\citealt{Blinnikov99}). As discussed
  above, observations indicate that this is not an important effect
  for GRB~061007.
\end{enumerate}

While all of these effects have the potential of softening the
low-energy slope from purely thermal emission, simulations are needed
in order to determine their relative importance. Indeed, recent
simulations by \cite{Mizuta10} show thermal emission from a GRB jet
with $\alpha = -0.5$, much softer than a blackbody spectrum with
$\alpha = 1$.

\subsection{Properties of the outflow}

Having identified a photospheric component in the spectrum we were
also able to derive the properties of the outflow, including the
Lorentz factor, the radius of the photosphere ($r_{\rm{ph}}$), the initial
radius of the outflow ($r_{\rm{0}}$) and the saturation radius
($r_{\rm{s}}$). As already discussed, the latter two parameters depend on
the assumption of non-dissipative fireball dynamics. We find that
$\Gamma$ varies between about 200 and $600~(\xi Y)^{1/4}$, while the
three radii remain relatively stable with average values of $r_{\rm{ph}} =
2.5 \pm 0.8 \times 10^{11}\ \xi^{-3/4}\ Y^{1/4}\ \mathrm{cm}$, $r_{\rm{0}}
= 3.5 \pm 1.7 \times 10^{8}\ \xi^{-4}\ Y^{-3/2}\ \mathrm{cm}$ and
$r_{\rm{s}} = 1.3 \pm 0.6 \times
10^{11}\ \xi^{-15/4}\ Y^{-5/4}\ \mathrm{cm}$. We especially note that
the values of the photospheric and saturation radii are rather similar
(see also Fig.~\ref{rphot}).  Since the radiative efficiency of the
thermal component scales as $\left( r_{\rm{ph}} / r_{\rm{s}} \right)^{-2/3}$
(\citealt{Meszaros00}) this implies a high efficiency, as expected
given that the photospheric component dominates the prompt emission.

As discussed in section \ref{photosphere}, the main uncertainty in
these numbers is that we do not know how far the power law extends
above the WAM energy band. In particular, if the power law extends to
GeV energies the values quoted above for $\Gamma$ and $r_{\rm{ph}}$ should
be about two times higher, while the the values for $r_{\rm{0}}$ and
$r_{\rm{s}}$ will be smaller. In addition, all the above parameters have a
weak dependence on the value of $Y$, which is the ratio of the energy
emitted in gamma rays to the total energy emitted by the burst. The
value of $Y$ can be measured through detailed afterglow observations
and is usually found to be less than 2-3 (e.g. \citealt{Cenko10}). In
the case of GRB~061007 it is hard to put a firm limit on $Y$ as there
is no jet break seen in the afterglow, but the preferred model of
\citet{Schady07} gives a value of around 10. As in the case of a power
law extending to high energies, this value would increase our
estimates of for $\Gamma$ and $r_{\rm{ph}}$ and decrease the values for
$r_{\rm{0}}$ and $r_{\rm{s}}$.

\subsection{The non-thermal component}

The power-law component in our model dominates at low and high
energies and is consistent with having the same slope ($s=-1.5$)
during the entire burst. Even though our energy range only extends up
to 5 MeV it is interesting to compare this component with the
high-energy power laws recently observed in some \fermi bursts. In
particular, we note that a power law slope of $-1.5$ has been observed
in several cases (\citealt{Guiriec10}) and that the isotropic energy
release of GRB 061007 is similar to the bright \fermi bursts in which
this component has been observed. We also note that the power-law
components that are required in addition to photospheric components in
BATSE GRBs have a preferred value close to $- 1.5$ (\citealt{Ryde06}).

There have been many suggestions for the origin of the extra
component, including external shocks (\citealt{Ghisellini10,Kumar10}),
hadronic processes (\citealt{Asano09,Razzaque10}), Compton
upscattering of a photospheric component (\citealt{Toma10}) as well as
a combination of different emission mechanisms (\citealt{Peer10b}). If
the peak of the prompt emission is identified with a photospheric
component it is also possible that the additional component is due to
synchrotron emission arising in the region above the photosphere. The
power-law slope of $s=-1.5$ supports this picture, as this is the
slope expected for fast-cooling synchrotron emission. However, in this
case the power law break must be above the \suzaku energy range, which
implies an unreasonably large $B$-field (for a synchrotron peak at
1~GeV and $\Gamma = 1000$ the magnetic field is $B \approx
10^{14}/\gamma_{m}^2\ \rm{G}$, where $\gamma_{m}^2$ is the minimum
Lorentz factor of the electrons).

Another option is that the power law is the inverse-Compton component
of low-energy synchrotron emission. In this picture the synchrotron
peak must be below the BAT energy range while the inverse Compton peak
is above the WAM energy band. Such a large separation between the two
peaks implies a very large efficiency for the electrons, which in turn
rules out the standard internal-shock scenario.

In order to evaluate models for the power-law component it is clearly
crucial to determine their breaks, something which is now becoming
possible with the \fermi satellite. So far, there is one clear
confirmation of a break in the case of GRB~090926A
(\citealt{Ackermann11}). For GRB~061007 we expect a break of the power
law in the LAT energy range (see section \ref{photosphere}), which
should encourage the search for breaks in similarly bright bursts in
the future.

\section*{Acknowledgments}

The authors thank Asaf Pe'er for useful comments on the manuscript.
JL acknowledges financial support from the Swedish Research Council
(VR) through the Oskar Klein Centre. FR acknowledges with thanks the
Swedish National Space Board for financial support.  This research was
partially supported by a Grant-in-Aid for Scientific Research
No. 19047001 (KY) of the Ministry of Education, Culture, Sports,
Science and Technology (MEXT).

\bibliographystyle{mn}
\bibliography{grbrefs}

\begin{thebibliography}{49}
\expandafter\ifx\csname natexlab\endcsname\relax\def\natexlab#1{#1}\fi

\bibitem[{{Abdo} {et~al.}(2009){Abdo}, {Ackermann}, {Ajello}, {Asano},
  {Atwood}, {Axelsson}, {Baldini}, {Ballet}, \& {et al.}}]{Abdo09}
{Abdo} A.~A., {et al.}, 2009, \apjl, 706, L138

\bibitem[{{Ackermann} {et~al.}(2011){Ackermann}, {Ajello}, {Baldini}, {Ballet},
  {Barbiellini}, {Bastieri}, {Bechtol}, {Bellazzini}, \& {et
  al.}}]{Ackermann11}
{Ackermann} M., {et al.}, 2011,  preprint (astro-ph/1101.2082)

\bibitem[{{Ackermann} {et~al.}(2010){Ackermann},  {Ajello}, {Asano}, {Axelsson}, 
	{Baldini}, {Ballet}, {Barbiellini}, {Baring}, \& {et al.}}]{Ackermann10}
{Ackermann} M., {et al.}, 2010, \apj, 716,
  1178

\bibitem[{{Asano} {et~al.}(2009){Asano}, {Guiriec}, \&
  {M{\'e}sz{\'a}ros}}]{Asano09}
{Asano} K., {Guiriec} S., {M{\'e}sz{\'a}ros} P., 2009, \apjl, 705, L191

\bibitem[{{Band} {et~al.}(1993){Band}, {Matteson}, {Ford}, {Schaefer},
  {Palmer}, {Teegarden}, {Cline}, {Briggs}, \& {et al.}}]{Band93}
{Band} D., {et al.}, 1993, \apj, 413, 281

\bibitem[{{Bellm}(2010)}]{Bellm10}
{Bellm} E.~C., 2010, \apj, 714, 881

\bibitem[{{Beloborodov}(2010)}]{Beloborodov09}
{Beloborodov} A.~M., 2010, \mnras, 407, 1033

\bibitem[{{Blinnikov} {et~al.}(1999){Blinnikov}, {Kozyreva}, \&
  {Panchenko}}]{Blinnikov99}
{Blinnikov} S.~I., {Kozyreva} A.~V., {Panchenko} I.~E., 1999, Astronomy
  Reports, 43, 739

\bibitem[{{Borgonovo} \& {Ryde}(2001)}]{Borgonovo01}
{Borgonovo} L., {Ryde} F., 2001, \apj, 548, 770

\bibitem[{{Cenko} {et~al.}(2010){Cenko}, {Frail}, {Harrison}, {Haislip},
  {Reichart}, {Butler}, {Cobb}, {Cucchiara}, \& {et al.}}]{Cenko10}
{Cenko} S.~B., {et al.}, 2010, preprint (astro-ph/1004.2900)

\bibitem[{{Crider} {et~al.}(1997){Crider}, {Liang}, {Smith}, {Preece},
  {Briggs}, {Pendleton}, {Paciesas}, {Band}, \& {et al.}}]{Crider97}
{Crider} A., {et al.}, 1997, \apjl,
  479, L39+

\bibitem[{{Edelson} \& {Krolik}(1988)}]{Edelson88}
{Edelson} R.~A., {Krolik} J.~H., 1988, \apj, 333, 646

\bibitem[{{Firmani} {et~al.}(2009){Firmani}, {Cabrera}, {Avila-Reese},
  {Ghisellini}, {Ghirlanda}, {Nava}, \& {Bosnjak}}]{Firmani09}
{Firmani} C., {Cabrera} J.~I., {Avila-Reese} V., {Ghisellini} G., {Ghirlanda}
  G., {Nava} L., {Bosnjak} Z., 2009, \mnras, 393, 1209

\bibitem[{{Ghirlanda} {et~al.}(2007){Ghirlanda}, {Bosnjak}, {Ghisellini},
  {Tavecchio}, \& {Firmani}}]{Ghirlanda07}
{Ghirlanda} G., {Bosnjak} Z., {Ghisellini} G., {Tavecchio} F., {Firmani} C.,
  2007, \mnras, 379, 73

\bibitem[{{Ghirlanda} {et~al.}(2003){Ghirlanda}, {Celotti}, \&
  {Ghisellini}}]{Ghirlanda03}
{Ghirlanda} G., {Celotti} A., {Ghisellini} G., 2003, \aap, 406, 879

\bibitem[{{Ghisellini} {et~al.}(2010){Ghisellini}, {Ghirlanda}, {Nava}, \&
  {Celotti}}]{Ghisellini10}
{Ghisellini} G., {Ghirlanda} G., {Nava} L., {Celotti} A., 2010, \mnras, 403,
  926

\bibitem[{{Golenetskii} {et~al.}(2006){Golenetskii}, {Aptekar}, {Mazets},
  {Pal'Shin}, {Frederiks}, \& {Cline}}]{Golenetskii06}
{Golenetskii} S., {Aptekar} R., {Mazets} E., {Pal'Shin} V., {Frederiks} D.,
  {Cline} T., 2006, GRB Coordinates Network, 5722, 1

\bibitem[{{Golenetskii} {et~al.}(1983){Golenetskii}, {Mazets}, {Aptekar}, \&
  {Ilinskii}}]{Golenetskii83}
{Golenetskii} S.~V., {Mazets} E.~P., {Aptekar} R.~L., {Ilinskii} V.~N., 1983,
  \nat, 306, 451

\bibitem[{{Granot} {et~al.}(2010){Granot}, {for the Fermi LAT Collaboration},
  \& {the GBM Collaboration}}]{Granot10}
{Granot} J., {for the Fermi LAT Collaboration}, {the GBM Collaboration}, 2010,
  preprint (astro-ph/1003.2452)

\bibitem[{{Guiriec} {et~al.}(2010{\natexlab{a}}){Guiriec}, {Briggs},
  {Connaugthon}, {Kara}, {Daigne}, {Kouveliotou}, {van der Horst}, {Paciesas},
  \& {et al.}}]{Guiriec10}
{Guiriec} S., {et al.},
  2010, \apj, 725, 225

\bibitem[{{Guiriec} {et~al.}(2010{\natexlab{b}}){Guiriec}, {Connaughton},
  {Briggs}, {Burgess}, {Ryde}, {Daigne}, {M{\'e}sz{\'a}ros}, {Goldstein}, \&
  {et al.}}]{Guiriec10b}
{Guiriec} S., {et al.},
  2010,  \apj, 727, L33

\bibitem[{{Jakobsson} {et~al.}(2006){Jakobsson}, {Fynbo}, {Tanvir}, \&
  {Rol}}]{Jakobsson06}
{Jakobsson} P., {Fynbo} J.~P.~U., {Tanvir} N., {Rol} E., 2006, GRB Coordinates
  Network, 5716, 1

\bibitem[{{Kumar} \& {Barniol Duran}(2010)}]{Kumar10}
{Kumar} P., {Barniol Duran} R., 2010, \mnras, 1243

\bibitem[{{Lazzati} \& {Begelman}(2010)}]{Lazzati10}
{Lazzati} D., {Begelman} M.~C., 2010, \apj, 725, 1137

\bibitem[{{Lu} {et~al.}(2010){Lu}, {Hou}, \& {Liang}}]{Lu10}
{Lu} R., {Hou} S., {Liang} E., 2010, \apj, 720, 1146

\bibitem[{{M{\'e}sz{\'a}ros} {et~al.}(2002){M{\'e}sz{\'a}ros}, {Ramirez-Ruiz},
  {Rees}, \& {Zhang}}]{Meszaros02}
{M{\'e}sz{\'a}ros} P., {Ramirez-Ruiz} E., {Rees} M.~J., {Zhang} B., 2002, \apj,
  578, 812

\bibitem[{{M{\'e}sz{\'a}ros} \& {Rees}(2000)}]{Meszaros00}
{M{\'e}sz{\'a}ros} P., {Rees} M.~J., 2000, \apj, 530, 292

\bibitem[{{Mizuta} {et~al.}(2010){Mizuta}, {Nagataki}, \& {Aoi}}]{Mizuta10}
{Mizuta} A., {Nagataki} S., {Aoi} J., 2010, preprint (astro-ph/1006.2440)

\bibitem[{{Mundell} {et~al.}(2007)}]{Mundell07}
{Mundell} C.~G., {et al.}, 2007, \apj, 660, 489

\bibitem[{{Ohno} {et~al.}(2009){Ohno}, {Ioka}, {Yamaoka}, {Tashiro},
  {Fukazawa}, \& {Nakagawa}}]{Ohno09}
{Ohno} M., {Ioka} K., {Yamaoka} K., {Tashiro} M., {Fukazawa} Y., {Nakagawa}
  Y.~E., 2009, \pasj, 61, 201

\bibitem[{{Osip} {et~al.}(2006){Osip}, {Chen}, \& {Prochaska}}]{Osip06}
{Osip} D., {Chen} H., {Prochaska} J.~X., 2006, GRB Coordinates Network, 5715, 1

\bibitem[{{Pe'er}(2008)}]{Peer08}
{Pe'er} A., 2008, \apj, 682, 463

\bibitem[{{Pe'er} {et~al.}(2006){Pe'er}, {M{\'e}sz{\'a}ros}, \&
  {Rees}}]{Peer06}
{Pe'er} A., {M{\'e}sz{\'a}ros} P., {Rees} M.~J., 2006, \apj, 642, 995

\bibitem[{{Pe'er} \& {Ryde}(2010)}]{Peer10}
{Pe'er} A., {Ryde} F., 2010, preprint (astro-ph/1008.4590)

\bibitem[{{Pe'er} {et~al.}(2007){Pe'er}, {Ryde}, {Wijers}, {M{\'e}sz{\'a}ros},
  \& {Rees}}]{Peer07}
{Pe'er} A., {Ryde} F., {Wijers} R.~A.~M.~J., {M{\'e}sz{\'a}ros} P., {Rees}
  M.~J., 2007, \apjl, 664, L1

\bibitem[{{Pe'er} {et~al.}(2010){Pe'er}, {Zhang}, {Ryde}, {McGlynn}, {Zhang},
  {Preece}, \& {Kouveliotou}}]{Peer10b}
{Pe'er} A., {Zhang} B., {Ryde} F., {McGlynn} S., {Zhang} B., {Preece} R.~D.,
  {Kouveliotou} C., 2010, preprint (astro-ph/1007.2228)

\bibitem[{{Peterson} {et~al.}(1998){Peterson}, {Wanders}, {Horne}, {Collier},
  {Alexander}, {Kaspi}, \& {Maoz}}]{Peterson98}
{Peterson} B.~M., {Wanders} I., {Horne} K., {Collier} S., {Alexander} T.,
  {Kaspi} S., {Maoz} D., 1998, \pasp, 110, 660

\bibitem[{{Preece} {et~al.}(1998){Preece}, {Briggs}, {Mallozzi}, {Pendleton},
  {Paciesas}, \& {Band}}]{Preece98}
{Preece} R.~D., {Briggs} M.~S., {Mallozzi} R.~S., {Pendleton} G.~N., {Paciesas}
  W.~S., {Band} D.~L., 1998, \apjl, 506, L23

\bibitem[{{Ramirez-Ruiz} \& {Lloyd-Ronning}(2002)}]{Ramirez-Ruiz02}
{Ramirez-Ruiz} E., {Lloyd-Ronning} N.~M., 2002, \na, 7, 197

\bibitem[{{Razzaque}(2010)}]{Razzaque10}
{Razzaque} S., 2010, \apjl, 724, L109

\bibitem[{{Rees} \& {M{\'e}sz{\'a}ros}(2005)}]{Rees05}
{Rees} M.~J., {M{\'e}sz{\'a}ros} P., 2005, \apj, 628, 847

\bibitem[{{Ryde}(2004)}]{Ryde04}
{Ryde} F., 2004, \apj, 614, 827

\bibitem[{{Ryde} {et~al.}(2010){Ryde}, {Axelsson}, {Zhang}, {McGlynn}, {Pe'er},
  {Lundman}, {Larsson}, {Battelino}, \& {et al.}}]{Ryde10}
{Ryde} F., 2010, \apjl, 709, L172

\bibitem[{{Ryde} {et~al.}(2006){Ryde}, {Bj{\"o}rnsson}, {Kaneko},
  {M{\'e}sz{\'a}ros}, {Preece}, \& {Battelino}}]{Ryde06}
{Ryde} F., {Bj{\"o}rnsson} C., {Kaneko} Y., {M{\'e}sz{\'a}ros} P., {Preece} R.,
  {Battelino} M., 2006, \apj, 652, 1400

\bibitem[{{Ryde} \& {Pe'er}(2009)}]{Ryde09}
{Ryde} F., {Pe'er} A., 2009, \apj, 702, 1211

\bibitem[{{Schady} {et~al.}(2007){Schady}, {de Pasquale}, {Page}, {Vetere},
  {Pandey}, {Wang}, {Cummings}, {Zhang}, \& {et al.}}]{Schady07}
{Schady} P., {et al.}, 2007, \mnras, 380, 1041

\bibitem[{{Toma} {et~al.}(2010){Toma}, {Wu}, \& {Meszaros}}]{Toma10}
{Toma} K., {Wu} X., {Meszaros} P., 2010, ArXiv e-prints

\bibitem[{{Yamaoka} {et~al.}(2006){Yamaoka}, {Sugita}, {Ohno}, {Takahashi},
  {Asano}, {Uehara}, {Fukazawa}, {Terada}, \& {et al.}}]{Yamaoka06}
{Yamaoka} K., {et al.}, 2006, GRB Coordinates Network, 5724, 1

\bibitem[{{Yonetoku} {et~al.}(2004){Yonetoku}, {Murakami}, {Nakamura},
  {Yamazaki}, {Inoue}, \& {Ioka}}]{Yonetoku04}
{Yonetoku} D., {Murakami} T., {Nakamura} T., {Yamazaki} R., {Inoue} A.~K.,
  {Ioka} K., 2004, \apj, 609, 935

\bibitem[{{Zhang} {et~al.}(2010){Zhang}, {Zhang}, {Liang}, {Fan}, {Wu},
  {Pe'er}, {Maxham}, {Gao}, \& {et al.}}]{Zhang10}
{Zhang} B., {et al.}, 2010, preprint (astro-ph/1009.3338)

\end{thebibliography}

\end{document}